\documentclass[preprintnumbers,amsmath,amssymbm,prd]{revtex4}
\usepackage{epsfig}
\usepackage{graphicx}

\begin{document}

\title{Analytic treatment of the excited instability spectra of the
magnetically charged SU(2) Reissner-Nordstr\"om black holes}
\author{Shahar Hod}
\address{The Ruppin Academic Center, Emeq Hefer 40250, Israel}
\address{ }
\address{The Hadassah Institute, Jerusalem 91010, Israel}
\date{\today}

\begin{abstract}
\ \ \ The magnetically charged SU(2) Reissner-Nordstr\"om black-hole
solutions of the coupled nonlinear Einstein-Yang-Mills field
equations are known to be characterized by infinite spectra of
unstable (imaginary) resonances
$\{\omega_n(r_+,r_-)\}_{n=0}^{n=\infty}$ (here $r_{\pm}$ are the
black-hole horizon radii). Based on direct {\it numerical}
computations of the black-hole instability spectra, it has recently
been observed that the excited instability eigenvalues of the
magnetically charged black holes exhibit a simple universal
behavior. In particular, it was shown that the numerically computed
instability eigenvalues of the magnetically charged black holes are
characterized by the small frequency universal relation
$\omega_n(r_+-r_-)=\lambda_n$, where $\{\lambda_n\}$ are
dimensionless constants which are independent of the black-hole
parameters. In the present paper we study analytically the
instability spectra of the magnetically charged SU(2)
Reissner-Nordstr\"om black holes. In particular, we provide a
rigorous {\it analytical} proof for the {\it numerically}-suggested
universal behavior $\omega_n(r_+-r_-)=\lambda_n$ in the small
frequency $\omega_n r_+\ll (r_+-r_-)/r_+$ regime. Interestingly, it
is shown that the excited black-hole resonances are characterized by
the simple universal relation
$\omega_{n+1}/\omega_n=e^{-2\pi/\sqrt{3}}$. Finally, we confirm our
analytical results for the black-hole instability spectra with
numerical computations.
\end{abstract}
\bigskip
\maketitle


\section{Introduction}

It is well known that the electrically charged U(1)
Reissner-Nordstr\"om black-hole spacetime describes a {\it stable}
solution of the coupled nonlinear Einstein-Maxwell field equations
\cite{Mon} (see also \cite{Hods}). On the other hand, the
magnetically charged SU(2) Reissner-Nordstr\"om black-hole spacetime
\cite{Yas} describes an {\it unstable} solution of the coupled
nonlinear Einstein-Yang-Mills field equations
\cite{Str,BizWal,Mas,Hodplb1}. In fact, the Reissner-Nordstr\"om
black-hole solutions of the Einstein-Yang-Mills theory are
characterized by an {\it infinite} spectrum of unstable perturbation
modes. These unstable (exponentially growing in time) modes are
described by an infinite set of imaginary black-hole resonances
$\{\omega_n\}_{n=0}^{n=\infty}$ \cite{Str,BizWal,Mas,Hodplb1}.

In a very interesting numerical investigation of the coupled
Einstein-Yang-Mills equations, it was recently revealed by Rinne
\cite{Oliv} that these unstable magnetically charged black-hole
solutions play the role of approximate \cite{Noteaprr}
codimension-two intermediate attractors (critical solutions
\cite{GunMar}) in the nonlinear gravitational collapse of the
Yang-Mills field \cite{Oliv,ChCh,ChHi,BizCha,Noteot}. In particular,
it has been shown explicitly \cite{Oliv} that the time spent in the
vicinity of an unstable magnetically charged SU(2)
Reissner-Nordstr\"om black-hole spacetime during a near-critical
gravitational collapse of the nonlinear Yang-Mills field can be
quantified by the characteristic scaling law \cite{Notepp}
\begin{equation}\label{Eq1}
\tau=\text{const}-\gamma\ln|p-p^*|\  .
\end{equation}

It is interesting to note that the critical exponents in the scaling
behavior (\ref{Eq1}), which characterizes the nonlinear
near-critical gravitational collapse of the Yang-Mills field, are
directly related to the imaginary eigenvalues which characterize the
instability spectrum of the corresponding magnetically charged
Reissner-Nordstr\"om black-hole spacetime \cite{Oliv}:
\begin{equation}\label{Eq2}
\gamma=1/|\omega_{\text{instability}}|\  .
\end{equation}
It is therefore physically interesting to investigate the
characteristic instability (imaginary) resonance spectra
$\{\omega_n(r_+,r_-)\}_{n=0}^{n=\infty}$ \cite{Noterpm} of these
magnetically charged black-hole solutions of the coupled nonlinear
Einstein-Yang-Mills equations.

In his important numerical work, Rinne \cite{Oliv} has recently
determined numerically the first three imaginary (unstable) resonant
frequencies which characterize the magnetically charged SU(2)
Reissner-Nordstr\"om black-hole spacetimes \cite{NoteRin}.
Subsequently, in \cite{HodEpjc} we have analyzed the detailed
numerical data provided by Rinne \cite{Oliv} and revealed the
intriguing fact that, to a good degree of accuracy, the numerically
computed \cite{Oliv} excited instability eigenvalues of the
magnetically charged SU(2) Reissner-Nordstr\"om black holes are
characterized by the remarkably simple {\it universal} behavior
\begin{equation}\label{Eq3}
\omega_n(r_+-r_-)=\lambda_n\ \ \ \ \text{for}\ \ \ \ \omega_n r_+\ll
(r_+-r_-)/r_+\  ,
\end{equation}
where $\{\lambda_n\}$ are dimensionless constants which seem to be
{\it independent} of the black-hole parameters.

The main goal of the present paper is to determine {\it
analytically} the characteristic instability spectra of the
magnetically charged SU(2) Reissner-Nordstr\"om black-hole solutions
of the coupled Einstein-Yang-Mills theory. In particular, in this
paper we shall provide a rigorous analytical proof for the validity
of the numerically suggested \cite{Oliv,HodEpjc} universal behavior
(\ref{Eq3}) which characterizes the excited instability spectra of
the SU(2) Reissner-Nordstr\"om black-hole spacetimes.

\section{Description of the system}

The SU(2) Reissner-Nordstr\"om black-hole spacetime of mass $M$ and
unit magnetic charge is characterized by the spherically-symmetric
line element \cite{Yas}
\begin{equation}\label{Eq4}
ds^2=-\Big(1-{{2m}\over{r}}\Big)dt^2+\Big(1-{{2m}\over{r}}\Big)^{-1}dr^2+r^2(d\theta^2+\sin^2\theta
d\phi^2)\ ,
\end{equation}
where the radially dependent mass function $m=m(r)$ is given by
\cite{Noteunit}
\begin{equation}\label{Eq5}
m(r)=M-{{1}\over{2r}}\  .
\end{equation}
The radii of the black-hole (outer and inner) horizons are given by
\begin{equation}\label{Eq6}
r_{\pm}=M\pm\sqrt{M^2-1}\  .
\end{equation}

As shown in \cite{Bizw}, linearized perturbation modes $\xi(r)
e^{-i\omega t}$ \cite{Notegro} of the magnetically charged
black-hole spacetime are governed by the wave equation
\begin{equation}\label{Eq7}
\Big[{{d^2}\over{dx^2}}+\omega^2-U(x) \Big]\xi=0\  ,
\end{equation}
where the radial coordinate $x=x(r)$ is defined by the differential
relation \cite{Notehor}
\begin{equation}\label{Eq8}
dx/dr=[1-2m(r)/r]^{-1}\  .
\end{equation}
The effective radial potential which governs the Schr\"odinger-like
wave equation (\ref{Eq7}) is given by \cite{Bizw}
\begin{equation}\label{Eq9}
U[x(r)]=-{1 \over {r^2}}\Big[1-{{2m(r)} \over r}\Big] \  .
\end{equation}

It is worth emphasizing the fact that the radial function $U(x)$ in
Eq. (\ref{Eq7}), which determines the spatial behavior of the
black-hole perturbation modes, has the form of an effective binding
potential. In particular, it is a negative definite function of the
radial coordinate $x$ and it vanishes asymptotically at the two
boundaries $x\to\pm\infty$ of the magnetically charged black-hole
spacetime. As shown in \cite{Bizw}, well-behaved perturbation modes
of the black-hole spacetime are characterized by spatially bounded
(exponentially decaying) radial eigenfunctions at the two asymptotic
boundaries:
\begin{equation}\label{Eq10}
\xi(x\to -\infty)\sim e^{|\omega|x}\to 0\
\end{equation}
and
\begin{equation}\label{Eq11}
\xi(x\to\infty)\sim xe^{-|\omega|x}\to 0\  ,
\end{equation}
where $\omega=i|\omega|$.

The radial differential equation (\ref{Eq7}), supplemented by the
physically motivated boundary conditions (\ref{Eq10}) and
(\ref{Eq11}) \cite{Bizw}, determine the discrete family
$\{\omega_n(r_+,r_-)\}_{n=0}^{n=\infty}$ of unstable ($\Im\omega>0$)
resonances which characterize the SU(2) Reissner-Nordstr\"om
black-hole spacetimes \cite{Mas,BizWal}. Interestingly, below we
shall show explicitly that the characteristic resonance spectrum of
the magnetically charged black holes can be studied analytically in
the regime $|\omega_n| r_+\ll1$ of small imaginary resonant
frequencies. In particular, we shall derive a remarkably compact
{\it analytical} formula [see Eq. (\ref{Eq31}) below] for the
excited instability eigenvalues which characterize the SU(2)
Reissner-Nordstr\"om black-hole solutions of the coupled
Einstein-Yang-Mills theory.

\section{The characteristic resonance condition}

The recent numerical results of Rinne \cite{Oliv} reveal that the
excited instability resonances $\{\omega_n\}_{n=1}^{n=\infty}$ of
the magnetically charged SU(2) Reissner-Nordstr\"om black-hole
spacetimes are characterized by the property
\begin{equation}\label{Eq12}
|\omega_n| r_+\ll1\ \ \ ; \ \ \ n=1,2,3,...
\end{equation}
As we shall now show, the Schr\"odinger-like differential equation
(\ref{Eq7}), which governs the dynamics of the black-hole
perturbation modes, is amenable to an {\it analytical} treatment in
the regime (\ref{Eq12}) of {\it small} resonant frequencies.

It was pointed out in \cite{Hodplb1} that, in the $M\gg1$ regime
(the regime of weakly-magnetized SU(2) Reissner-Nordstr\"om black
holes), the Schr\"odinger-like perturbation equation (\ref{Eq7}) can
be transformed using the well-known Chandrasekhar transformations
\cite{Chan} to the physically equivalent Teukolsky-like radial
equation \cite{TeuPre}:
\begin{equation}\label{Eq13}
\Delta^2{{d^2\psi}\over{dr^2}}+\Big[{{\omega^2r^4+2iM\omega
r^2}}-\Delta[2i\omega r+\ell(\ell+1)]\Big]\psi=0\ ,
\end{equation}
where the complex number
\begin{equation}\label{Eq14}
\ell\equiv {{-1+i\sqrt{3}}\over{2}}\
\end{equation}
plays the role of an effective spherical harmonic index (see
\cite{Hodplb1} for details), and \cite{TeuPre}
\begin{equation}\label{Eq15}
\Delta(r;M\gg1)=r^2-2Mr\  .
\end{equation}
The mathematical Chandrasekhar transformations \cite{Chan} can also
be used in the case of generic magnetically charged SU(2)
Reissner-Nordstr\"om black holes \cite{Notegen}, in which case the
generalized expression for the radial function $\Delta(r)$ in the
Teukolsky-like radial perturbation equation (\ref{Eq13}) is given by
\cite{TeuPre}
\begin{equation}\label{Eq16}
\Delta(r;M)=r^2-2Mr+1\  .
\end{equation}
It is convenient to use the dimensionless physical variables
\cite{Page,Hodcen}
\begin{equation}\label{Eq17}
z\equiv {{r-r_+}\over {r_+-r_-}}\ \ \ ; \ \ \ k\equiv
-i\omega(r_+-r_-)\ \ \ ; \ \ \ \varpi\equiv
-i\omega{{2Mr_+}\over{r_+-r_-}}\ ,
\end{equation}
in terms of which the radial differential equation (\ref{Eq13})
reads
\begin{eqnarray}\label{Eq18}
z^2(z+1)^2{{d^2\psi}\over{dz^2}}
+\big[-k^2z^4+2kz^3-\ell(\ell+1)z(z+1)-\varpi(2z+1)-\varpi^2\big]\psi&=&0\
.
\end{eqnarray}

The physically acceptable solution \cite{Noteph1} of the radial
perturbation equation (\ref{Eq18}) in the near-horizon $kz\ll 1$
region \cite{Notenear} can be expressed in terms of the familiar
hypergeometric function \cite{Page,Hodcen,Abram}:
\begin{eqnarray}\label{Eq19}
\psi(z)=z^{1+\varpi}(z+1)^{1-\varpi}
{_2F_1}(-\ell+1,\ell+2;2+2\varpi;-z)\  .
\end{eqnarray}
The physically acceptable solution \cite{Noteph2} of the radial
perturbation equation (\ref{Eq18}) in the asymptotic region $z\gg
\varpi+1$ \cite{Notefar} can be expressed in terms of the familiar
confluent hypergeometric function \cite{Page,Hodcen,Abram}:
\begin{eqnarray}\label{Eq20}
\psi(z)=Ae^{kz}z^{\ell+1}{_1F_1}(\ell+2;2\ell+2;-2kz)
+Be^{kz}z^{-\ell}{_1F_1}(-\ell+1;-2\ell;-2kz)\ ,
\end{eqnarray}
For small resonant frequencies in the regime (\ref{Eq12}), the
values of the dimensionless coefficients $A$ and $B$ in (\ref{Eq20})
can be determined by matching the two solutions [(\ref{Eq19}) and
(\ref{Eq20})] of the radial perturbation equation (\ref{Eq18}) in
the overlap region \cite{Noteov}
\begin{equation}\label{Eq21}
\varpi+1\ll z\ll 1/k\  .
\end{equation}
This matching procedure yields the expressions \cite{Hodcen,Page}
\begin{equation}\label{Eq22}
A={{\Gamma(2\ell+1)\Gamma(2+2\varpi)}\over
{\Gamma(\ell+2)\Gamma(\ell+1+2\varpi)}}\
\end{equation}
and
\begin{equation}\label{Eq23}
B={{\Gamma(-2\ell-1)\Gamma(2+2\varpi)}\over
{\Gamma(-\ell+1)\Gamma(-\ell+2\varpi)}}\
\end{equation}
for the dimensionless coefficients $\{A,B\}$ in (\ref{Eq20}).

Substituting the expressions (\ref{Eq22}) and (\ref{Eq23}) into the
far-region expression (\ref{Eq20}) of the radial eigenfunction and
using the asymptotic ($z\gg1$) properties of the confluent
hypergeometric functions \cite{Abram}, one finds the large-$z$
asymptotic behavior \cite{Hodcen,Page}
\begin{eqnarray}\label{Eq24}
\psi(z\to\infty)=\psi_1 re^{-kz}+\psi_2 r^{-1}e^{kz}\
\end{eqnarray}
of the radial eigenfunctions, where
\begin{eqnarray}\label{Eq25}
(r_+-r_-)\psi_1={{(2\ell+1)\Gamma^2(2\ell+1)\Gamma(2+2\varpi)}\over{\Gamma^2(\ell+2)
\Gamma(\ell+1+2\varpi)}}(-2k)^{-\ell}
-{{(2\ell+1)\Gamma^2(-2\ell-1)\Gamma(2+2\varpi)}\over{\Gamma^2(-\ell+1)\Gamma(-\ell+2\varpi)}}(-2k)^{\ell+1}\
,
\end{eqnarray}
and
\begin{eqnarray}\label{Eq26}
(r_+-r_-)^{-1}\psi_2={{(2\ell+1)\Gamma^2(2\ell+1)\Gamma(2+2\varpi)}\over{\ell(\ell+1)\Gamma^2(\ell)
\Gamma(\ell+1+2\varpi)}}(2k)^{-\ell-2}
-{{(2\ell+1)(\ell+1)\Gamma^2(-2\ell-1)\Gamma(2+2\varpi)}\over{\ell\Gamma^2(-\ell)\Gamma(-\ell+2\varpi)}}(2k)^{\ell-1}\
.
\end{eqnarray}

A spatially bounded (normalizable) radial eigenfunction which
satisfies the physically motivated boundary condition (\ref{Eq11})
at spatial infinity \cite{Bizw} is characterized by the asymptotic
relation $\psi(z\to\infty)\to 0$. Thus, the coefficient $\psi_2$ of
the exploding exponent in the asymptotic expression (\ref{Eq24})
should vanish, yielding the characteristic resonance equation [see
Eq. (\ref{Eq26})]
\begin{eqnarray}\label{Eq27}
(2k)^{2\ell+1}=\Big[{{\Gamma(2\ell+1)\Gamma(-\ell)}\over{(\ell+1)\Gamma(-2\ell-1)\Gamma(\ell)}}\Big]^2
{{\Gamma(-\ell+2\varpi)}\over{\Gamma(\ell+1+2\varpi)}}\ \nonumber
\\
\end{eqnarray}
for the instability eigenvalues of the SU(2) Reissner-Nordstr\"om
black-hole spacetimes. Taking cognizance of Eq. (\ref{Eq14}), one
can write the resonance equation (\ref{Eq27}) for the instability
spectra of the magnetically charged black holes in the form
\cite{Notegam}
\begin{eqnarray}\label{Eq28}
k^{i\sqrt{3}}=8^{i\sqrt{3}}e^{i2\pi/3}{{\Gamma^2(i{{\sqrt{3}}\over{2}})}
\over{\Gamma^2(-i{{\sqrt{3}}\over{2}})}}
{{\Gamma({{1-i\sqrt{3}}\over{2}}+2\varpi)}\over{\Gamma({{1+i\sqrt{3}}\over{2}}+2\varpi)}}\
 .
\end{eqnarray}

\section{The black-hole excited instability spectra}

As we shall now show, the characteristic resonance equation
(\ref{Eq28}) for the instability eigenvalues of the magnetically
charged SU(2) Reissner-Nordstr\"om black holes can be solved
analytically in the regime \cite{Notewrpm}
\begin{equation}\label{Eq29}
\varpi\ll1\  .
\end{equation}
In this small frequency regime the resonance equation (\ref{Eq28})
can be approximated by
\begin{eqnarray}\label{Eq30}
k^{i\sqrt{3}}=8^{i\sqrt{3}}e^{i2\pi/3}{{\Gamma^2(i{{\sqrt{3}}\over{2}})}
\over{\Gamma^2(-i{{\sqrt{3}}\over{2}})}}
{{\Gamma({{1-i\sqrt{3}}\over{2}})}\over{\Gamma({{1+i\sqrt{3}}\over{2}})}}\
,
\end{eqnarray}
which yields the characteristic {\it infinite} spectrum
\cite{Notenn,Notecu}
\begin{eqnarray}\label{Eq31}
\omega_n\times
(r_+-r_-)=i\times8e^{-{{2\pi}\over{\sqrt{3}}}(n-{1\over3})+{{4\theta-2\phi}\over{\sqrt{3}}}}\
\ ; \ \ n=1,2,3,...\
\end{eqnarray}
of unstable ($\Im\omega>0$) black-hole resonances, where
\begin{equation}\label{Eq32}
\theta\equiv\arg[\Gamma({{i\sqrt{3}}/{2}})]\ \ \ ; \ \ \
\phi\equiv\arg[\Gamma({{1+i\sqrt{3}}\over{2}})]\  .
\end{equation}

It is worth emphasizing again that the analytically derived formula
(\ref{Eq31}) for the characteristic instability spectra of the
magnetically charged SU(2) Reissner-Nordstr\"om black holes is valid
in the small frequency regime [see (\ref{Eq17}) and (\ref{Eq29})]
\begin{equation}\label{Eq33}
\omega_n r_+\ll {{r_+-r_-}\over{r_+}}\  .
\end{equation}
This inequality implies that, for a given value of the black-hole
dimensionless temperature $(r_+-r_-)/r_+$, the analytical formula
(\ref{Eq31}) describes an infinite family of unstable (imaginary)
black-hole resonances in the regime
\begin{equation}\label{Eq34}
n\gtrsim |\ln({{r_+-r_-}\over{r_+}})|+1\  .
\end{equation}

It is interesting to note that the instability spectra (\ref{Eq31})
of the magnetically charged SU(2) Reissner-Nordstr\"om black-hole
spacetimes have the simple generic form $\omega_n\times
(r_+-r_-)=\text{constant}_n\equiv \lambda_n$ [see Eq. (\ref{Eq3})].
We have therefore provided here an {\it analytical} proof for the
{\it numerically}-observed \cite{Oliv,HodEpjc} universal behavior
(\ref{Eq3}) of the black-hole excited instability eigenvalues.

\section{Numerical confirmation}

It is of considerable physical interest to verify the validity of
the analytically derived formula (\ref{Eq31}) for the excited
instability eigenvalues of the SU(2) Reissner-Nordstr\"om black
holes. The corresponding instability eigenvalues of the magnetically
charged black holes were recently computed numerically in the very
interesting work of Rinne \cite{Oliv}. In Table I we present the
dimensionless ratio $\omega^{\text{ana}}_2/\omega^{\text{num}}_2$,
where $\{\omega^{\text{ana}}_2(r_+)\}$ are the analytically
calculated excited instability eigenvalues of the SU(2)
Reissner-Nordstr\"om black holes as given by the analytical formula
(\ref{Eq31}) and $\{\omega^{\text{num}}_2(r_+)\}$ are the
numerically computed \cite{Oliv} instability eigenvalues of the
magnetically charged black holes. From the data presented in Table I
one finds a fairly good agreement between the analytically derived
formula (\ref{Eq31}) for the excited instability spectra of the
SU(2) Reissner-Nordstr\"om black holes and the corresponding
numerically computed black-hole instability eigenvalues.

\begin{table}[htbp]
\centering
\begin{tabular}{|c|c|c|c|c|c|c|}
\hline
\ \ $r_+$\ \ & \ 10.0\ \ & \ \ 8.0\ \ & \ \ 6.0\ \ & \ \ 4.0\ \ & \ \ 2.0\ \ & \ 1.5\ \ \\
\hline \ \ $\omega^{\text{ana}}_2/\omega^{\text{num}}_2$\ \ & \ \
1.046\ \ \ & \ \ 1.025\ \ \ & \ \ 1.011\ \ \ &
\ \ 1.005\ \ \ & \ \ 1.007\ \ \ & \ \ 1.012\ \ \ \\
\hline
\end{tabular}
\caption{The excited instability eigenvalues of the magnetically
charged SU(2) Reissner-Nordstr\"om black holes. We display the
dimensionless ratio $\omega^{\text{ana}}_2/\omega^{\text{num}}_2$,
where $\{\omega^{\text{ana}}_2(r_+)\}$ are the analytically
calculated instability eigenvalues of the SU(2) Reissner-Nordstr\"om
black holes as given by the analytical formula (\ref{Eq31}) and
$\{\omega^{\text{num}}_2(r_+)\}$ are the corresponding numerically
computed \cite{Oliv} instability eigenvalues of the magnetically
charged black holes. One finds a fairly good agreement between the
numerical data of \cite{Oliv} and the analytically derived formula
(\ref{Eq31}) for the excited instability spectra of the SU(2)
Reissner-Nordstr\"om black holes.} \label{Table1}
\end{table}

\section{Summary}

The magnetically charged SU(2) Reissner-Nordstr\"om black-hole
spacetimes describe a family of unstable solutions of the coupled
nonlinear Einstein-Yang-Mills field equations
\cite{Str,BizWal,Hodplb1,Oliv}. In particular, these magnetically
charged black holes are known to be characterized by {\it infinite}
spectra of imaginary (unstable) resonant frequencies
$\{\omega_n(r_+,r_-)\}_{n=0}^{n=\infty}$. Based on direct {\it
numerical} computations of the black-hole instability spectra
\cite{Oliv}, it has recently been pointed out \cite{HodEpjc} that
the excited instability eigenvalues of the magnetically charged
SU(2) Reissner-Nordstr\"om black holes are described, to a very good
degree of accuracy, by the simple universal relation (\ref{Eq3}).

In the present paper we have studied analytically the characteristic
instability spectra of the magnetically charged SU(2)
Reissner-Nordstr\"om black-hole spacetimes in the small frequency
regime. In particular, we have provided a simple {\it analytical}
proof for the {\it numerically}-observed \cite{Oliv,HodEpjc}
universal behavior [see Eq. (\ref{Eq3})]
\begin{equation}\label{Eq35}
\omega_n(r_+-r_-)=\lambda_n\ \ \ \ \text{for}\ \ \ \ \omega_n r_+\ll
(r_+-r_-)/r_+\
\end{equation}
which characterizes the black-hole excited instability resonances,
where $\{\lambda_n\}$ are constants. Our analysis has revealed that
these dimensionless constants (which are {\it independent} of the
black-hole parameters) are given by the simple relation [see Eqs.
(\ref{Eq31}) and (\ref{Eq32})]
\begin{equation}\label{Eq36}
\lambda_n=i\times8e^{-{{2\pi}\over{\sqrt{3}}}(n-{1\over3})+{{4\theta-2\phi}\over{\sqrt{3}}}}\
.
\end{equation}

Finally, it is interesting to note that one finds from (\ref{Eq31})
the simple dimensionless ratio
\begin{equation}\label{Eq37}
{{\omega_{n+1}}\over{\omega_n}}=e^{-2\pi/\sqrt{3}}\
\end{equation}
for the characteristic excited instability eigenvalues of the
magnetically charged SU(2) Reissner-Nordstr\"om black holes. It is
worth emphasizing the fact that the relation (\ref{Eq37}), which
characterizes the instability resonance spectra of the magnetically
charged black-hole spacetimes, is {\it universal} in the sense that
it is {\it independent} of the physical parameters (masses and
magnetic charges) of the SU(2) Reissner-Nordstr\"om black holes.

\bigskip
\noindent {\bf ACKNOWLEDGMENTS}

This research is supported by the Carmel Science Foundation. I would
like to thank Oliver Rinne for sharing with me his numerical data. I
would also like to thank Yael Oren, Arbel M. Ongo, Ayelet B. Lata,
and Alona B. Tea for helpful discussions.


\bigskip

\begin{thebibliography}{99}

\bibitem{Mon} V. Moncrief, Phys. Rev. D {\bf 9}, 2707 (1974); V. Moncrief, Phys.
Rev. D {\bf 10}, 1057 (1974).

\bibitem{Hods} S. Hod, Phys. Lett. B {\bf 713}, 505 (2012);
S. Hod, Phys. Lett. B {\bf 718}, 1489 (2013) [arXiv:1304.6474]; S.
Hod, Phys. Rev. D {\bf 91}, 044047 (2015) [arXiv:1504.00009].

\bibitem{Yas} P. B. Yasskin, Phys. Rev. D {\bf 12}, 108 (1975).

\bibitem{Str} N. Straumann and Z.-H. Zhou, Phys. Lett. B {\bf 237}, 353 (1990);
N. Straumann and Z.-H. Zhou, Phys. Lett. B {\bf 243}, 33 (1990).

\bibitem{BizWal} P. Bizo\'n and R. M. Wald, Phys. Lett. B {\bf 267}, 173 (1991).

\bibitem{Mas} P. Breitenlohner, P. Forg\'acs, and D. Maison, Nucl. Phys. B {\bf 383},
357 (1992); P. Breitenlohner, P. Forg\'acs, and D. Maison, Nucl.
Phys. B {\bf 442}, 126 (1995).

\bibitem{Hodplb1} S. Hod, Phys. Lett. B {\bf 739}, 157 (2014)
[arXiv:1410.7406]; S. Hod, Phys. Lett. B {\bf 661}, 175 (2008)
[arXiv:0803.0608].

\bibitem{Oliv} O. Rinne, Phys. Rev. D {\bf 90}, 124084 (2014).

\bibitem{Noteaprr} As discussed in \cite{Oliv}, the SU(2) Reissner-Nordstr\"om
black-hole spacetime is only an approximate intermediate attractor
of the nonlinear gravitational collapse of the Yang-Mills field
because it is characterized by an {\it infinite} family of
exponentially growing (unstable) perturbation modes.

\bibitem{GunMar} See C. Gundlach and J. M. Mart\'in-Garc\'ia,
Living Rev. Relativity {\bf 10} (2007), for an excellent review on
the black-hole critical phenomena in nonlinear gravitational
collapse.

\bibitem{ChCh} M. W. Choptuik, T. Chmaj, and P. Bizo\'n, Phys. Rev. Lett. {\bf 77}, 424 (1996);
C. Gundlach, Phys. Rev. D {\bf 55}, 6002 (1997).

\bibitem{ChHi} M. Choptuik, E. Hirschmann, and R. Marsa, Phys. Rev. D {\bf 60}, 124011
(1999).

\bibitem{BizCha} P. Bizo\'n and T. Chmaj, Phys. Rev. D {\bf 61}, 067501 (2000).

\bibitem{Noteot} It is worth noting that
this physically interesting fact refers to type I and Type III
nonlinear critical behaviors in gravitational collapse, see
\cite{ChCh,ChHi,BizCha,Oliv} for details.

\bibitem{Notepp} Here the quantity $|p-p^*|$ provides a measure in the physical parameter space
for the distance of the initial field data from the threshold
(critical) solution of the nonlinear Einstein-Yang-Mills theory
\cite{GunMar}.

\bibitem{Noterpm} Here $r_{\pm}$ are the black-hole horizon radii [see Eq. (\ref{Eq6}) below].

\bibitem{NoteRin} As noted above, the SU(2) Reissner-Nordstr\"om black-hole solutions of
the coupled nonlinear Einstein-Yang-Mills field equations are
characterized by {\it infinite} spectra
$\{\omega_n(r_+,r_-)\}_{n=0}^{n=\infty}$ of imaginary (unstable)
resonant frequencies \cite{Mas}. Reference \cite{Oliv} has provided,
for the first time, detailed numerical results for the first three
resonant frequencies which quantify the instability growth rates of
these magnetically charged black-hole spacetimes.

\bibitem{HodEpjc} S. Hod, The Euro. Phys. Journal C {\bf 75}, 180 (2015) [arXiv:1504.04010].

\bibitem{Noteunit} We shall use natural units in which $G=c=\hbar=1$.

\bibitem{Bizw} P. Bizo\'n, Phys. Lett B {\bf 259}, 53 (1991).

\bibitem{Notegro} Note that unstable (exponentially growing in time) perturbation modes of
the magnetically charged black-hole spacetime are characterized by
the relation $\Im\omega>0$.

\bibitem{Notehor} Note that the near-horizon region $r \to r_+$ of the black-hole spacetime is mapped by
the differential relation (\ref{Eq8}) to $x \to -\infty$, whereas
spatial infinity $r\to\infty$ is mapped to $x\to\infty$.

\bibitem{Chan} S. Chandrasekhar, Proc. R. Soc. London {\bf A 343}, 289 (1975).

\bibitem{TeuPre} S. A. Teukolsky and W. H. Press, Astrophys. J. {\bf 193}, 443
(1974).

\bibitem{Notegen} We use here the term `generic SU(2) Reissner-Nordstr\"om black holes' to
describe magnetically charged black holes with generic masses (that
is, in the present analysis we do {\it not} assume the strong
inequality $M\gg1$ previously assumed in \cite{Hodplb1}).

\bibitem{Page} D. N. Page, Phys. Rev. D {\bf 13}, 198 (1976).

\bibitem{Hodcen} S. Hod, Phys. Lett. B {\bf 666}, 483 (2008)
[arXiv:0810.5419]; S. Hod, Phys. Rev. D {\bf 88}, 084018 (2013)
[arXiv:1311.3007].

\bibitem{Noteph1} That is, the mathematical solution which respects the physically motivated near-horizon
boundary condition (\ref{Eq10}).

\bibitem{Notenear} Note that in the near-horizon region $kz\ll 1$ one can
neglect the first two terms inside the square brackets in Eq.
(\ref{Eq18}) \cite{Page}.

\bibitem{Abram} M. Abramowitz and I. A. Stegun, {\it Handbook of
Mathematical Functions} (Dover Publications, New York, 1970).

\bibitem{Noteph2} That is, the mathematical solution which respects the physically motivated asymptotic boundary condition
(\ref{Eq11}) at spatial infinity.

\bibitem{Notefar} Note that in the asymptotic region $z\gg \varpi+1$ one can
neglect the last two terms inside the square brackets in Eq.
(\ref{Eq18}) \cite{Page}.

\bibitem{Noteov} It is worth noting that the overlap radial region $\varpi\ll z\ll
1/k$ exists in the regime $|\omega| r_+\ll1$ of small resonant
frequencies [see Eqs. (\ref{Eq12}) and (\ref{Eq17})].

\bibitem{Notegam} Here we have used Eq. 6.1.18 of \cite{Abram}.

\bibitem{Notewrpm} Note that this regime corresponds to $\omega r_+\ll
(r_+-r_-)/r_+$ [see Eq. (\ref{Eq17})].

\bibitem{Notenn} Here we have used the relation $1=e^{-i2\pi n}$, where the integer
$n=1,2,3,...\ $ is the resonance parameter of the black-hole
perturbation mode. In addition, we have used here Eq. 6.1.23 of
\cite{Abram}.

\bibitem{Notecu} As noted in \cite{Hodplb1}, one finds the numerical ratio
$\theta/2\phi=1.0016$. Thus, one can replace, with an accuracy of
$0.05\%$, the expression $({{4\theta-2\phi})/{\sqrt{3}}}$ in
(\ref{Eq31}) by the simpler term $\sqrt{3}\theta$.

\end{thebibliography}
\end{document}